\documentclass{article}

\usepackage{arxiv}

\usepackage[utf8]{inputenc} 
\usepackage[T1]{fontenc}    
\usepackage{hyperref}       
\usepackage{url}            
\usepackage{booktabs}       
\usepackage{amsfonts}       
\usepackage{nicefrac}       
\usepackage{microtype}      
\usepackage{lipsum}
\usepackage{graphicx}
\usepackage{float}
\graphicspath{ {./images/} }

\title{Modular Fault-Tolerant DBMS}

\author{
 Fot Nikolay \\
  Institute of Mathematics and Information Technologies\\
  Orenburg State University\\
  Orenburg, Russia, 460052 \\
  \texttt{fot\_ns@office.osu.ru} \\
   \And
 Vinarsky Alexander \\
  Department of Physical and Quantum Electronics\\
  Moscow Institute of Physics and Technology\\
  Dolgoprudny, Russia, 141701 \\
  \texttt{vinarskii.aa@phystech.edu} \\
}

\begin{document}
\maketitle
\begin{abstract}
The article addresses the problem of storing data in extreme environmental conditions with limited computing resources and memory. There is a requirement to create portable, fault-tolerant, modular database management systems (DBMS) that are optimized for use in embedded systems. Existing databases, such as LittleDB, LMDB, and Berkeley DB, are reviewed, and their limitations are identified. A variant of a portable DBMS is introduced to efficiently manage data in environments where computational resource usage must be minimized, while meeting specific requirements for fault tolerance and noise immunity. Common solutions for optimizing of insertion, storage and management of data are reviewed. Algorithms for fault-tolerant data encoding in RAM are implemented. An architectural solution to data storage and minimizing the impact of bit errors is proposed. Software that manages relational data in extreme conditions is developed, that allows testing and comparing results with existing solutions.
\end{abstract}

\section{Introduction}

\par Modern database management systems (DBMS) are universal solutions for management of large amount of information, nonetheless, the functionality provided by DBMS in some cases is excessive, resulting in technical requirements that exceed the capabilities of existing hardware platforms.
\par But with the growth of technological level, problems appear, so storage and computation in enviroments where full-fledged personal computers and large disk space cannot be used, are required to solve them. For instance, applications such as satellites and autonomous rovers.

\par Existing DBMS include:
\begin{sloppypar}
    \begin{enumerate} 
        \item LittleDB [2] - An SQL database developed by Pouria Moosavi, that is suitable for embedded systems (basic automation systems). It uses a relational model.
        \item LMDB [3] - A DBMS developed by Symas company, deployable on basic automation controllers. It is based on the key-value architecture.
        \item Berkeley DB [4] - A DBMS developed by Oracle Corporation, that uses key-value structure. Like other lightweight DBMS, it can be deployed on basic automation controllers with integrated standard libraries.
    \end{enumerate}
\end{sloppypar}

\par One of the major disadvantages of the above-mentioned solutions is the lack of a mechanism of data protection from single-event upsets (SEUs), While this issue may be partially addressed through the use of radiation-resistant memory with ECC mechanisms [5], nonetheless, it does not resolve the major problem, namely the lack of noise immunity of data stored in read-only memory (ROM).
\par Hence, the development of small-sized modular DBMS, that allows managing data effectively in difficult conditions and is able to operate under constraints such as limited disk space and processing power, is relevant. Such DBMS can be implemented for basic automation in systems using low-cost microcontrollers (e.g., STM32F103C6T8 [6]) or higher-cost, noise-immune microcontrollers like the 1874BE7T [7] and 1887BE6T [8].


\section{Features of the developed architecture}

\par Since similar DBMS require an extensive amount of random access memory (RAM), in the development of modular lightweight DBMS it was not feasible to use the established architectural solutions [9], resulting in development of memory manager, the principle of which is managing of static fixed-size memory allocation via blocks, based on the mechanism of single-linked list.
\par The memory manager and standard libraries for interacting with it are integrated into the DBMS kernel, which comprises:

\begin{enumerate}
    \item Input tokenizer - Handles command splitting into separate directives.
    \item Syntax parser - Handles directive execution with tokenized parameters.
    \item Table abstraction layer, which implements management of directory layer with support of parallel computations and of creation of data processing threads.
    \item Directory abstraction layer, which implements management of page layer with support of parallel computations and of creation of data processing threads.
    \item Page layer - The deepest abstraction layer. Pages in the DBMS kernel function similarly to paging in the Linux kernel, storing data and error-correcting bits and processing them via Hamming code for bit error correction [10].
\end{enumerate}

\par In contrast to DBMS solutions that store data in one file, there is a data fragmentation system in the developed software, which allows storing data in large amounts of fixed-size blocks, which are distributed over directories of the file system. The implementation of this approach has made it possible to achieve following properties:

\begin{enumerate}
    \item The increase of resistance to single-event upsets (SEUs) of the entire system due to physical data decompression.
    \item Delegation of the tasks of finding pages and directories to the file system.
    \item Optimization of algorithms for deleting and adding data without requiring additional memory.
\end{enumerate}

\par The last offset for insertion is saved in the developed software, which allows to avoid iteration through allocated memory while searching for free space. This design optimizes the process of adding pages, directories, and tables. At the same time, the deletion operation shifts this offset to the position of the last data erasure, allowing to avoid excessive memory usage.
\par Moreover, in order to minimize disk space accesses, which are computationally expensive in terms of performance and platform resources, a system of global cache table was integrated into the kernel, the main purpose of which is storing tables, directories and pages in random access memory (RAM) (utilizing a data eviction algorithm that factors in their relevance, lock state, and memory cleanup requirements). When the system attempts to insert a new entry, the table algorithm first attempts to write it into available free space. If this fails, it initiates an attempt to rewrite older entries based on their lock status. If insertion is still impossible,  the entry is temporarily held in a dedicated memory buffer, which is freed immediately after the entry is no longer needed.


\section{DBMS Performance Analysis and Testing}

\begin{sloppypar}
\par To assess the quality of the developed software, a set of metrics were defined:
\par For the memory footprint metric, executable files of the same project compiled in different systems were analyzed. The results are summarized in Table 1.

\end{sloppypar}

\begin{table}[H]
 \caption{Executable size on different platforms}
  \centering
  \begin{tabular}{lll}
    \toprule
    \cmidrule(r){1-2}
    Platform     & Dynamic size     & Static size (B) \\
    \midrule
    Debian GNU/Linux 11 & 32121 & 111992 \\
    Fedora 40 & 35536 & 99480 \\
    5.15.146.1-microsoft-standart-WSL & 37376 & 91632 \\
    MacOSX 10.15.7 & 86624 & - \\
    \bottomrule
  \end{tabular}
  \label{tab:table}
\end{table}

\par The dynamic size represents the total project size, whereas the static size reflects the executable size on embedded systems. Thus, the minimum requirement for deployment is at least 120 KB of flash memory.

\begin{sloppypar}
\par For comparison, Table 2 lists the sizes of lightweight executable versions of various DBMS.

\end{sloppypar}

\begin{table}[H]
 \caption{Executable size of different DBMS on GNU Linux Ubuntu}
  \centering
  \begin{tabular}{lll}
    \toprule
    \cmidrule(r){1-2}
    DBMS     & Static size (B) \\
    \midrule
    Developed DBMS & 111992 \\
    LittleDB & 174173 \\
    SQLLite & 1476288 \\
    Greenplum [12] & $\approx35000000$ \\
    PostgreSQL 12 [13] & $\approx40000000$ \\
    Oracle DB [14] & $\approx2339017432$ \\
    \bottomrule
  \end{tabular}
  \label{tab:table}
\end{table}

\par To assess the performance of the developed DBMS, a test setup was deployed on a Raspberry Pi 3B platform equipped with a BCM2835 processor, 909 MB of RAM, and a 32 GB memory card.
\par The test data comprised a 36-character string containing a record index, a random string variable, and a random integer variable. The results are presented in Table 3.

\begin{table}[H]
 \caption{DBMS average work time}
  \centering
  \begin{tabular}{lll}
    \toprule
    \cmidrule(r){1-2}
    Records count     & Insert (sec.) & Record get (sec.) \\
    \midrule
    100000 & 196.5696 & 0.034014 \\
    1000000 & 1902.4908 & 0.034913 \\
    5000000 & 9422.5949 & 0.041493 \\
    \bottomrule
  \end{tabular}
  \label{tab:table}
\end{table}

\par For comparison with these results, the results of analogous desktop DBMS are presented in Table 4.

\begin{table}[H]
 \caption{Insert speed for 5 mil. records}
  \centering
  \begin{tabular}{lll}
    \toprule
    \cmidrule(r){1-2}
    Records count     & Insert (k. sec.) \\
    \midrule
    Developed DBMS & 9.422 \\
    PostgreSQL 12 & 198 \\
    Oracle & 216 \\
    SQL Server & 312 \\
    Greenplum & 660 \\
    \bottomrule
  \end{tabular}
  \label{tab:table}
\end{table}


\section{Summary}

\par Thus, the developed modular DBMS demonstrates high performance under constrained computational resources. Due to its multi-level architecture with parallel access support, mechanism of caching, and integration of bit error correction algorithm, the system ensures resilience to single-event upsets (SEUs) and minimizes disk space accesses.
\par The software solution  significantly enhances the efficiency of autonomous rovers, embedded systems, and similar resource-constrained devices.

\newpage

\bibliographystyle{unsrt}  

\end{document}